\documentclass[prd,aps,twocolumn,groupedaddress,showpacs]{revtex4}
\begin{document}
\input{psfig.sty}

\title{Accuracy requirements for the calculation of gravitational
waveforms from coalescing compact binaries in numerical relativity}

\author{Mark Miller}
\affiliation{Jet Propulsion Laboratory, California Institute of Technology,
Pasadena, California 91109, USA}

\date{\today}

\begin{abstract}

I discuss the accuracy requirements on numerical relativity calculations
of inspiraling compact object binaries whose extracted 
gravitational waveforms are to be used as templates 
for matched filtering signal extraction and physical parameter estimation in 
modern interferometric gravitational wave detectors.  Using a post-Newtonian
point particle model for the pre-merger phase of the binary inspiral,
I calculate the maximum allowable errors for the mass and 
relative velocity and positions of the binary
during numerical simulations of the binary inspiral.
These maximum allowable errors
are compared to the errors of
state-of-the-art numerical simulations of multiple-orbit
binary neutron star calculations in full general relativity, and are found
to be smaller by several orders of magnitude.  A post-Newtonian
model for the error of these numerical simulations suggests that 
adaptive mesh refinement coupled with second order accurate finite
difference codes will {\it not} be able to robustly obtain the accuracy 
required for reliable gravitational wave extraction on Terabyte-scale
computers.  I conclude that higher order methods (higher order finite
difference methods and/or spectral methods) combined with adaptive 
mesh refinement and/or multipatch technology will be needed 
for robustly accurate gravitational wave extraction from numerical 
relativity calculations of binary coalescence scenarios.

\end{abstract}


\maketitle

\section{Introduction}

Studies in numerical relativity in the past decades have
claimed at least partial motivation from the imminent direct 
detection of gravitational waves from both ground based 
(LIGO, GEO, TAMA, VIRGO) and space based (LISA) detectors.  Theoretical
calculations of gravitational waveforms from realistic astrophysical
phenomena will be an essential ingredient in the extraction of 
characteristic information of gravitational wave sources (e.g.,
mass, spin, size, and composition of compact objects) from the detected
gravitational waves.  In particular, 
gravitational waves produced during
the coalescence of binary compact objects (neutron stars and/or black holes)
are strong candidates for direct detection, and thus it is precisely these
systems that are of great interest to the numerical relativity community.
Recent advances in numerical relativity, in particular with respect to
the stability of binary black hole 
evolutions~\cite{Bruegmann:2003aw,
Alcubierre2003:pre-ISCO-coalescence-times,Brandt00} and
binary neutron star evolutions~\cite{Miller:2003vc,Shibata02a,Marronetti04}, 
make possible the
calculation of gravitational waves from fully general and consistent
numerical relativity simulations of binary coalescences.   

However, numerical relativity simulations contain errors that
arise from attempting to solve continuum differential equations (the
Einstein field equations) on infinite domains (asymptotically flat 
spacetimes) with digital computers of finite size and speed.
Examples
of these errors
are truncation errors (e.g., due to the truncating of Taylor 
series expansions
for finite difference methods or to the truncating of
function expansions for spectral methods) and boundary errors
(e.g., errors induced by 
the introduction of a computational domain in causal contact with the binary
and/or the gravitational waves being emitted).
The magnitude of these errors determines the accuracy of the numerical
relativity simulation (here, I do not include modeling errors in
the determination of the accuracy of a numerical relativity code, e.g., 
inaccurate equation of state for neutron star matter or
astrophysically incorrect initial data for binary simulations).
In this paper, I demonstrate, for the first time, a calculation for 
determining the accuracy required of inspiraling binary numerical
relativity simulations in order that the characteristics of the 
extracted gravitational waveform represent the physics of the binary
system to the experimental error level of the gravitational wave detector.
Using the gravitational waveform accuracy criterion 
in~\cite{Flanagan97b},
I calculate the sensitivity of the gravitational waveform
to various parameters of the dynamical binary system (e.g., binary separation, 
angular velocity, mass) assuming a specific target sensitivity for the
gravitational wave detector; forcing the errors in the same
dynamical parameters within numerical relativity coalescence simulations
to be smaller than these sensitivities will be one way of guaranteeing
an extracted theoretical
waveform accurate to the sensitivity level of the gravitational 
wave detector.
Truncation and boundary
errors in the orbital separation of the multiple-orbit binary neutron
star simulations in~\cite{Miller:2003vc} are calculated and are shown
to be several orders of magnitude {\it larger} than the margin allowed for
by the gravitational wave sensitivity calculation.  Using a post-Newtonian
model of the truncation and boundary errors in the binary neutron star
numerical relativity simulations, I estimate the computational 
resources required for accurate gravitational waveform generation from
such simulations, and conclude that both higher order methods and
mesh refinement or multipatch technology will be required for
robust, reliable, and accurate gravitational waveform extraction from
numerical relativity simulations of coalescing binary inspirals and mergers.

The outline of the rest of the paper is as follows.
In Section~\ref{sec:accuracy}, the
sensitivity of gravitational waveforms to various physical characteristics
of the coalescing binary is calculated for binary black hole and
binary neutron star systems.  In Section~\ref{sec:nsaccuracy}, I compare
these sensitivities to errors in recent 
state-of-the-art binary simulations in numerical
relativity, and find the errors to be orders of magnitude 
larger than the gravitational
wave sensitivities.  I conclude by discussing methods that may
be helpful in reducing the error of numerical relativity calculations
of coalescing binaries to levels that would permit their extracted
gravitational waveforms to be used with confidence 
as templates in modern interferometric
gravitational wave detectors.

\section{Required accuracy for numerical relativity simulations of 
binary inspirals:  sensitivity of gravitational waveforms to physical
characteristics of the binary}
\label{sec:accuracy}

In order to calculate the sensitivity of gravitational waveforms to various
physical characteristics of the coalescing binary, I calculate binary
coalescence solutions to
the post-Newtonian equations
of motion for non-spinning point particles.  This formalism has the advantage
that accurate gravitational waveforms can be calculated in astrophysically
relevant binary coalescence scenarios.  The major disadvantage is that
at small binary separations, the point particle approximation breaks down
due to finite size effects;
the gravitational waveforms obtained by solving the 
post-Newtonian equations of motion must therefore be truncated at the
point where these effects become important.
As a result, the sensitivities calculated 
here will only bound the true sensitivity, i.e. the
sensitivity of the entire gravitational wavetrain, through plunge, merger, and
ringdown of the final merged object.  
In other words, due to the fact that the gravitational waveform is being
truncated when finite size effects become important, the sensitivity of
the entire physical gravitational waveform to 
variations in the physical characteristics
of the system is being {\it underestimated}.
Thus, the errors in numerical relativity simulations must
be {\it at least as small as} the sensitivities
calculated here.

\subsection{Post-Newtonian equations of motion}
\label{sec:pneom}

The general relativistic equations of motion
for non-spinning point particles in harmonic coordinates
with positions $\vec{x}_1$ and
$\vec{x}_2$, and masses $m_1$ and $m_2$, can be written in a post-Newtonian
expansion as
\begin{eqnarray}
\frac {d^2 \vec{x}} {dt^2} & = & - \frac {m}{r^2} \hat{n}  +
   \frac {m}{r^2} \left [ \hat{n} (A_{1PN} + A_{2PN} +
                                   A_{3PN} + \cdots) \right. + \nonumber \\
   & & \left.  \dot{r} \vec{v} (B_{1PN} + B_{2PN} + B_{3PN} + \cdots
   ) \right ] + \nonumber \\
   & & \frac {8}{5} \eta \frac {m}{r^2} \frac {m}{r} \left [ \dot{r}
      \hat{n} (A_{2.5PN} + A_{3.5PN} + A_{4.5PN} + \cdots ) -
      \right. \nonumber \\
   & & \left. \vec{v} (B_{2.5PN} + B_{3.5PN} + B_{4.5PN} + \cdots )
      \right ],
\label{eq:pneom}
\end{eqnarray}
where $\vec{x} = \vec{x}_2 - \vec{x}_1$ is the relative separation of
the particles, $\vec{v} = \vec{v}_2 - \vec{v}_1$ is the relative
velocity between the particles, $r = |\vec{x}|$, $\hat{n} = \vec{x}/r$,
$m = m_1 + m_2$, $\eta = m_1 m_2 / m^2$, and $\dot{r} = dr/dt$.
The post-Newtonian expansion in Eq.~\ref{eq:pneom}
is carried out in powers
of $\epsilon \sim m/r \sim v^2$ (I set $G=c=1$),
where each power of $\epsilon$ represents one
post-Newtonian (PN) order in the series.
The 1PN and 2PN terms are standard (e.g., see~\cite{Damour81,Itoh01,Pati02}).
The 2.5PN~\cite{Blanchet98,Itoh01,Pati02} and 3.5PN~\cite{Pati02} have
also been completely determined.  The 3.0PN terms have just recently
been calculated~\cite{Blanchet03} up to one gauge-dependent constant.
Employing an energy and angular momentum
balance technique, the 4.5PN terms have been fixed modulo
$12$ free ``gauge'' parameters~\cite{Gopakumar97} 
(appendix B of~\cite{Miller03c} demonstrates that these free parameters have
a negligible effect on inspiral dynamics).  Once the initial relative 
positions and velocities of the
particles are given, Eq.~\ref{eq:pneom} is then solved numerically
for the time evolution of the binary system.  Specifically, if
the initial separation $r$ and its time derivative $\dot{r}$ along
with the initial relative angular position $\phi$ and its time derivative
$\dot{\phi}$ are specified at time $t=0$, then the equations of motion
Eq.~\ref{eq:pneom} specifies $r(t)$ and $\phi(t)$ for $t > 0$
(I assume the binary orbits within the $z=0$ plane).

I use the post-Newtonian formalism presented
in~\cite{Epstein75,Wagoner76,Turner78,Lincoln90} to calculate the 
polarization state $h(t) = h_{+}(t)$ of the
gravitational radiation as a function of the motion of the binary.
For definiteness, I fix both observation angles $(\Phi,\Theta)$
to be $0$ for the remainder of the paper (i.e., the
binary, which is orbiting in the $z=0$ plane, is observed along the $+z$-axis).

\subsection{Waveform sensitivity to physical characteristics of the binary}

An inner
product on the space of waveforms $h(t)$ is defined as
\begin{equation}
\left \langle h_1 | h_2 \right \rangle = 4 \: \verb+Re+ \: 
   \left \{ \int_0^\infty df \:
   \frac {{\tilde{h}^{*}_1}(f) {\tilde{h}_2}(f)}{S_h(f)} \right \},
\end{equation}
where ${\tilde{h}_1}(f)$ and ${\tilde{h}_2}(f)$ are the Fourier 
transforms of the two waveforms $h_1(t)$ and $h_2(t)$, and
$S_h(f)$ is the one-sided power spectral density of the strain noise
of the detector. 
For the calculations in the remainder of this paper, the
model of the one-sided power spectral density of the strain 
noise for the advanced LIGO detector found in~\cite{Flanagan97a}
is used for $S_h(f)$, where the mass scale is set by assuming an equal mass
binary with $m_1 = m_2 = 1.4 \: M_{\odot}$.  The latest detection
rates for binary neutron star coalescences for the advanced LIGO
sensitivity is between $40$ and $650$ events per
year~\cite{Kalogera04}.
Assume that the
gravitational waveform $h(t)$ 
contains an error $\delta h(t)$.
Arguments in~\cite{Flanagan97b}
determine the criteria that the error $\delta h(t)$ 
be small enough so that the quantity
\begin{equation}
\Delta \equiv \frac {1}{2} 
\frac { \left \langle \delta h | \delta h \right \rangle}
      { \left \langle h | h \right \rangle }
\label{eq:delta}
\end{equation}
satisfies
\begin{equation}
\Delta \leq 0.01 .
\label{eq:delta100th}
\end{equation}
This accuracy criteria is based on matched filtering accuracy arguments,
physical parameter estimation arguments, and arguments based on the total
information content of the gravitational wave signal 
(see \cite{Flanagan97b} for details).   
The specific criteria Eq.~\ref{eq:delta100th}
is based on the assumption that the signal to noise ratio
is of order $10$.  Much higher signal to noise ratios are expected for LISA,
and therefore an even more stringent requirement on $\delta h(t)$ could
be required in that case 
(e.g., for gravitational wave templates used in physical parameter
estimation, the required $\Delta$ scales like the inverse square of
the signal to noise ratio).

Using the post-Newtonian equations of motion described
in Sec.~\ref{sec:pneom} as a model for coalescing binary compact objects,
along with the gravitational wave accuracy requirements 
Eqs.~\ref{eq:delta}~and~\ref{eq:delta100th},
I now analyze the sensitivity of the gravitational waveform to various
physical characteristics of the binary dynamics.  Note that the gravitational
waveform ${h}(t)$ of our post-Newtonian model of a binary inspiral 
depends solely on the $8$ parameters $\vec{\lambda} = \{ r_0, \phi_0, 
\dot{r}_0, \dot{\phi}_0, m_1, m_2, t_f, R \}$.  Here,
$r_0$ and $\phi_0$ specifies the relative binary position at the initial time
$t = t_0$, $\dot{r}_0$ and $\dot{\phi}_0$ specifies the initial 
relative binary velocity, $m_1$ and $m_2$ specifies the mass of each
compact object, while $R$ specifies the observation
distance from the center of mass of the binary.  The
parameter $t_f$ is the time at which the waveform is truncated
due to the breakdown of the post-Newtonian point particle approximation.  
This breakdown occurs when the internal structure of
the individual
compact objects has a significant effect on
the orbital dynamics of the binary.  Throughout the rest of this paper,
I take $t_f$ to be the time at which the binary separation $r(t_f) = 4 \: m$
for equal mass binary black holes and $r(t_f) = 8 \: m$ for equal mass
binary neutron stars.  I orient the system such that the initial 
relative angle $\phi_0 = 0$.  For a specific initial binary separation
$r_0$, I set the initial relative velocity parameters $\dot{r}_0$ and
$\dot{\phi}_0$ to be those specified by the unique {\it quasicircular solution}
of the binary (the quasicircular solution to the PN equations of motion
is obtained by starting from circular orbit initial data in the limit
as the initial separation $r \rightarrow \infty$, see
Sec.~2 
of~\cite{Miller03c}).

For a specific set of parameters
$\vec{\lambda}$ that specify the waveform $h(t)$, 
define the quantity
$\Delta_{0.01}\vec{\lambda}$ to be the ranges in the parameters
such 
that the change in the gravitational waveform $\delta h(t)$ induced
by separately changing each individual component $\lambda_i$ 
satisfies $\Delta \leq 0.01$.
The quantity $\Delta_{0.01}\vec{\lambda}$
can be interpreted as the ``allowable'' error in each parameter during
the course of a numerical relativity time evolution simulation, since
changes within this range do not appreciably (to the tolerance 
set by Eq.~\ref{eq:delta}) affect the gravitational waveform. 

\begin{figure}
\vspace{0.0cm}
\hspace{0.0cm}
\psfig{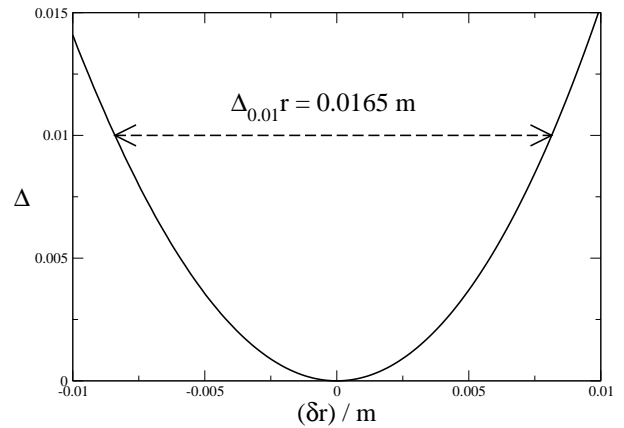}
\caption{An example of the definition of the $\Delta_{0.01}$ operator.
$\Delta$ (Eq.~\ref{eq:delta}) is plotted as a function of the variation
of the initial orbital separation $\delta r$ for an equal-mass black hole
binary, starting $3$ orbits before merger ($m$ is the total mass of the 
system).  In order that variations in the resulting gravitational waveform
satisfy $\Delta < 0.01$, any variation in the initial orbital separation
$r$ must satisfy $\delta r < 0.0165 \: m$.   I therefore define
$\Delta_{0.01}r = 0.0165 \: m$.
}
\vspace{0.0cm}
\label{fig:delta}
\end{figure}

As a concrete example, I examine the sensitivity of the gravitational
waveform on changes in the binary separation $r$.  I solve the 
4.5PN equations of motion for the quasicircular solution for an equal
mass ($m_1 = m_2 = m/2$) binary and find that exactly 3 orbits before
merger (for definiteness, assume binary black holes
and define the merger to be at
binary separation $r = 4 \: m$), the binary separation is $r = 7.3195 \: m$,
the relative radial velocity is $\dot{r} = -0.005573$, and the 
relative angular velocity is $\dot{\phi} = 0.04263 / m$.  Using these
values of parameters as initial data, the subsequent gravitational waveform
$h(t)$ for the last 3 orbits before merger is computed.
I define a second gravitational waveform ${h}^{\prime}(t)$ to
be the one obtained by changing the initial binary separation
$r_0 \rightarrow r_0 + \delta r$, keeping all other initial parameters
fixed.
The change in the waveform $\delta h(t)$ induced by changing the 
initial binary separation an amount $\delta r$ is therefore
$\delta h(t) = {h}^{\prime}(t) - h(t)$.
The quantity $\Delta$ (Eq.~\ref{eq:delta}) can now be calculated; 
in Fig.~\ref{fig:delta}, the quantity $\Delta$
is plotted as a function of the change in initial binary separation
$\delta r$.  The quantity ${\Delta}_{0.01} r$ is defined as the 
range of $\delta r$ such that $\Delta \leq 0.01$;  in this case,
${\Delta}_{0.01} r = 0.0165 \: m$ (see Fig.~\ref{fig:delta}).
The intuitive interpretation of this calculations is as follows:
$\delta r$
represents the error in the binary separation of
a numerical relativity simulation of
equal mass black holes at a time when 3 orbits remain until the
merger.  The quantity ${\Delta}_{0.01} r$ represents the 
``allowable'' error in binary separation $r$ as set by the 
tolerance level of Eq.~\ref{eq:delta100th}.  Thus, a numerical error
$| \delta r |$ in the binary separation of a numerical relativity simulation
at a time when 3 orbits remain until merger that is 
greater than ${\Delta}_{0.01} r$ would correspond to an 
unacceptable level of error in that simulation.

\begin{figure}
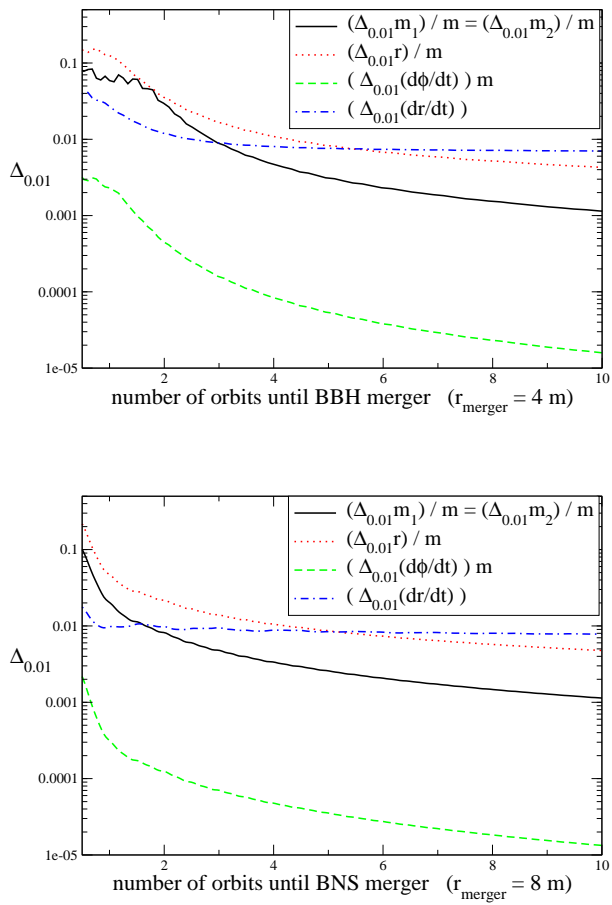

\mbox{}
\vspace{0.0cm}
\hspace{0.0cm}
\psfig{figure=fig2a.eps,width=8cm}
\mbox{}
\vspace{1.0cm}
\hspace{0.0cm}
\psfig{figure=fig2b.eps,width=8cm}
\caption{
Gravitational wave sensitivities to various dynamical quantities for orbiting 
binary black holes (top panel) and binary neutron stars (lower panel) are 
plotted as a function of the number of orbits remaining until merger.
Shown are gravitational wave sensitivities of the mass of each 
compact object ($\Delta_{0.01} m_1~=~\Delta_{0.01} m_2$, solid line),
the orbital separation ($\Delta_{0.01}r$, dotted line), 
the relative orbital angular velocity ($\Delta_{0.01}\dot{\phi}$, dashed line),
and the relative orbital radial velocity ($\Delta_{0.01}\dot{r}$, alternating
dot-dashed line).
}
\vspace{0.0cm}
\label{fig:bbhsensitivity}
\end{figure}

In Fig.~\ref{fig:bbhsensitivity},
the parameter tolerance $\Delta_{0.01} \lambda_i$
for physical parameters $m_1$, $m_2$, $r_0$, 
$\dot{\phi}_0$ and $\dot{r}_0$ is plotted as a function of the number of
orbits until merger for equal mass binary black holes and
binary neutron stars.  Note that the
gravitational waveform is roughly an order of magnitude more sensitive 
to small changes in the angular velocity as compared to small 
changes in the radial velocity.

\section{Case study: binary neutron star evolution}
\label{sec:nsaccuracy}

\subsection{comparing errors in numerical relativity simulations
of coalescing binary neutron stars to gravitational waveform 
sensitivities}
\label{sec:ns1}

Several recent studies~\cite{Miller:2003vc,Marronetti04} have analyzed various 
aspects of orbiting binary neutron stars by performing 3+1 general
relativistic hydrodynamic simulations.  It is instructive to 
analyze the accuracy of these simulations in comparison to the 
sensitivity required for the accurate generation of gravitational waves
as represented in Fig.~\ref{fig:bbhsensitivity}.  

In~\cite{Miller:2003vc}, quasiequilibrium initial 
data sets corresponding to two 
equal mass,
corotating binary neutron stars were numerically evolved using 
a general 3+1 numerical relativity code that simultaneously
solved the Einstein field equations coupled to the general relativistic
hydrodynamics equations.  In order to assess the quality of the numerical
solution, the exact same initial data set was numerically evolved
5 separate times, using a variety of discretization parameters and 
outer boundary placements (see Table II in~\cite{Miller:2003vc}).
There are two types of numerical errors associated with the 
simulations presented in~\cite{Miller:2003vc}.  The first type of
numerical error is due to
the finite difference approximation, where derivatives in the 
partial differential equations have been replaced by truncated
Taylor-series approximations.  Assuming that the finite difference
equations used in~\cite{Miller:2003vc} are consistent and stable, Lax 
convergence theorems (see~\cite{Gustafsson95}) state that the 
relationship between any quantity $Q$ constructed 
from a true solution
of the differential equation is related to that same quantity observed in
the numerical solution $Q_{numerical}$ by
\begin{equation}
Q = Q_{numerical} + C_1 (\Delta x) + C_2 {(\Delta x)}^2 + \cdots,
\end{equation}
where $C_i$ are constants (which are different for distinct quantities $Q$)
and $\Delta x$ is the discretization parameter used in the construction of
the finite difference equations.  Assuming that
$\Delta x$ is made small enough so that higher order terms can be ignored, 
define the truncation error ${(\Delta Q)}_{trunc}$
of the calculations in~\cite{Miller:2003vc}
to be
\begin{equation}
{(\Delta Q)}_{trunc} = C_1 (\Delta x) + C_2 {(\Delta x)}^2.
\label{eq:truncerror}
\end{equation}
While the code being analyzed is formally second order 
convergent in both space and time, the use of high resolution
shock capturing (HRSC) methods renders the hydrodynamics convergence
rate to be first order in space in regions where the dynamical variables
obtain a local extrema (see, e.g.,~\cite{Hirsch92}).  Thus, the 
form of the truncation error, Eq.~\ref{eq:truncerror}, necessarily 
contains a term proportional to the first power of the spatial
discretization parameter $\Delta x$.

The second type of numerical error that exists in the simulations 
presented in~\cite{Miller:2003vc} is due to the location of the 
boundary of the computational domain.  Ideally, the computational domain
would be placed many gravitational wavelengths away from center of
mass of the orbiting binary, thus minimizing the effect of the boundary
on the dynamics of the binary.  However, computational resource limitations
coupled with the fact that the code used in~\cite{Miller:2003vc} 
is only second order accurate and does
not employ adaptive mesh refinement, prevented the location of the 
outer boundary of the computational domain to be placed no farther than
$1/4$ of a gravitational wavelength from the center of mass of the 
system.  I model the error associated with the close proximity of 
the boundary of the computational domain analogously with that of 
the truncation error:  assume that the error in any particular
quantity $Q$ induced by the 
boundary goes to $0$ as the distance between the center of mass of
the system and the location of the outer boundary (denote this
distance as $r_b$) goes to infinity.  
I expand this boundary-induced
error, ${(\Delta Q)}_{bound}$, as a power series about
$r_b = \infty$ and assume that the values of $r_b$ used in the calculations
in~\cite{Miller:2003vc} are large enough so that 
the power series can be truncated as
\begin{equation}
{(\Delta Q)}_{bound} = \frac {B_1} { r_b }  + \frac {B_2} { {r_b}^2 }
\end{equation}
The total numerical error, therefore, is represented as the relationship
between the exact quantity $Q_{exact}$ 
specified from the solution to the differential
equations and quantity $Q_{numerical}$ produced by numerical simulations:
\begin{equation}
Q_{exact} = Q_{numerical} + {(\Delta Q)}_{trunc} + {(\Delta Q)}_{bound}.
\label{eq:qerror}
\end{equation}

\begin{figure}
\vspace{0.0cm}
\hspace{0.0cm}
\psfig{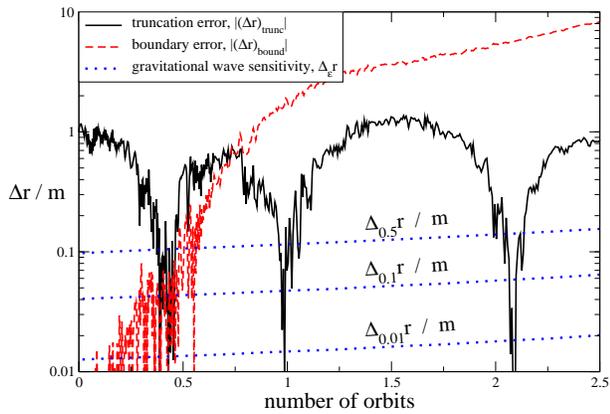}
\caption{The absolute value of the truncation error ${(\Delta r)}_{trunc}$
and boundary error ${(\Delta r)}_{bound}$ of the orbital
separation is plotted as a function
of the number of orbits for the binary neutron star simulations 
in~\cite{Miller:2003vc}.  For comparison, the gravitational wave
sensitivity to variations in the orbital separation, $\Delta_\epsilon r$
is also plotted.  Assuming the use of gravitational waveforms produced
from numerical simulations as matched filtering
search templates in a gravitational wave detector, sensitivities of
$\Delta_{0.01}$, $\Delta_{0.1}$, and $\Delta_{0.5}$
correspond to a detector event loss rate of 
$3\%$, $27\%$, and $87\%$, respectively.
}
\vspace{0.0cm}
\label{fig:varns}
\end{figure}

I now compare the errors in the binary separation $r$ contained in 
the simulations of~\cite{Miller:2003vc} with the sensitivity
of the gravitational waveform to changes in the binary
separation, $\Delta_{0.01} r$.
I use Eq.~\ref{eq:qerror} to calculate the numerical errors
in the calculation of the binary separation $r(t)$ for the simulations
presented in~\cite{Miller:2003vc}.  At each time $t$, I assume that
the numerically computed binary separation $r_{numerical}$ takes the
form of Eq.~\ref{eq:qerror}:
\begin{equation}
r_{exact} = r_{numerical} + {(\Delta r)}_{trunc} + {(\Delta r)}_{bound}.
\label{eq:rerror}
\end{equation}
Using the results from the 
five simulations displayed in Fig.~17~in~\cite{Miller:2003vc}
(which are reproduced in Fig.~\ref{fig:numrelpncomp} of this paper), the 
five unknowns in Eq.~\ref{eq:rerror} ($r_{exact}$, $C_1$, $C_2$,
$B_1$, and $B_2$) are specified at each time $t$.  The 
absolute values of the truncation error
${(\Delta r)}_{trunc}$ and the boundary error ${(\Delta r)}_{bound}$
of the binary separation are plotted in Fig.~\ref{fig:varns}.  For
comparison, the gravitational wave sensitivity to binary 
separation, $\Delta_{0.01} r$ 
(see Fig.~\ref{fig:bbhsensitivity}), is also plotted.  Important to
notice is the magnitude of the errors (both truncation errors
and boundary errors) in the binary separation of the numerical relativity
calculations in~\cite{Miller:2003vc} as compared to the sensitivity of
the gravitational waveform to the binary separation:  the 
simulation errors in the binary separation are 
several orders of magnitude larger than the 
gravitational waveform sensitivity to variations in the binary separation!
Assuming that they are to be used as signal detection and parameter estimation 
templates, the gravitational waveforms extracted from these numerical
simulations will contain errors that dominate the experimental errors in
modern interferometric gravitational wave detectors.
Steps must be taken to
improve the accuracy of these simulations before gravitational
waves extracted from them can be considered for use as templates
in gravitational wave detectors.

\subsection{estimating the accuracy and computational resources
required to extract sufficiently accurate waveforms from binary 
coalescence simulations}
\label{sec:ns2}

\begin{figure}
\vspace{0.0cm}
\hspace{0.0cm}
\psfig{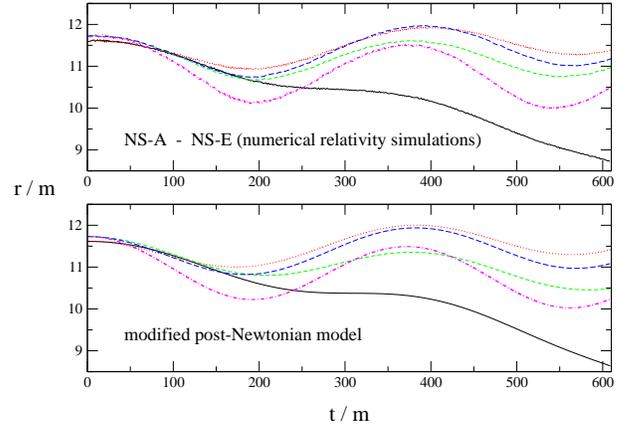}
\caption{Top panel:  plotted is the
orbital separation as a function of time from 
the numerical relativity simulations of orbiting binary neutron stars
in~\cite{Miller:2003vc}.  These simulations were performed with the
same initial data, but with different
discretization parameters $\Delta x$ and computational domain boundary 
placement
parameters $r_b$.  Shown are simulations NS-A (solid line), NS-B (dotted line),
NS-C (short dashed line), NS-D (long dashed line), and NS-E (alternating
dot-dashed line), evolved to time $t = 610 \: m$, which corresponds to
$2.5$ orbits.  Bottom panel: the modified post-Newtonian model
$r_{\tt{pn}}(t,{(\Delta x)}_j,{(r_b)}_j,{\sigma}^{min}_i)$ is plotted for the 
five discretization and boundary parameters $\{ {(\Delta x)}_j,{(r_b)}_j \}$, 
$(j=1,2,3,4,5)$, corresponding to the parameters used in
numerical relativity simulations
NS-A through NS-E, respectively.  The modified post-Newtonian model
robustly encapsulates the effects of the truncation and boundary 
errors within the full numerical relativity simulations.
}
\vspace{0.0cm}
\label{fig:numrelpncomp}
\end{figure}

In order to obtain information on the computational resources required to
reduce the boundary and truncation errors of simulations such as those
in~\cite{Miller:2003vc} down to acceptable levels (i.e., such that
the errors in the gravitational waveforms $\delta h$ induced by these
numerical errors satisfy $\Delta \leq 0.01$), these errors
and their effect on gravitational waveforms are modeled
using the post-Newtonian
equations of motion for spinless point particles, Eq.~\ref{eq:pneom}.
The same mass and initial orbital separation is used in the post-Newtonian
model as was used in the numerical
relativity simulations.  The initial angular velocity of the binary is
determined by the circular orbit assumption (see~\cite{Miller03c}), which
is consistent with the initial data used for the numerical relativity
simulations in~\cite{Miller:2003vc}.
In order to model the effects of truncation and boundary errors with
the post-Newtonian model, it is modified in the following way.
First, the initial angular velocity of the binary is set to be 
a function of discretization $\Delta x$ and outer boundary 
placement $r_b$ as
\begin{eqnarray}
{\dot{\phi}}_{0}(\Delta x,r_b) & =  {\dot{\phi}}_{circular} + &
\sigma_1 {(\frac {\Delta x}{m})} + 
\sigma_2 {(\frac {\Delta x}{m})}^2  \nonumber \\ 
 & & + \sigma_3 {(\frac {m}{r_b})} +
       \sigma_4 {(\frac {m}{r_b})}^2 
\label{eq:phidotmodel}
\end{eqnarray}
where ${\dot{\phi}}_{circular}$ is the angular velocity determined by
the circular orbit assumption.  Second, the post-Newtonian equations
of motion are modified so that the evolution of the angular 
momentum is given by
\begin{eqnarray}
{\left ( \frac {dL}{dt} \right )}_{} &  = &
{\left ( \frac {dL}{dt} \right )}_{PN} + 
\sigma_5 {(\frac {\Delta x}{m})} +
\sigma_6 {(\frac {\Delta x}{m})}^2  \nonumber \\ 
 & & \hspace{2.0cm} + \sigma_7 {(\frac {m}{r_b})} +
       \sigma_8 {(\frac {m}{r_b})}^2 
\label{eq:ldotmodel}
\end{eqnarray}
In the ideal limit of infinite resolution 
($\Delta x \rightarrow 0$) and infinite distance from the center of
mass of the system to the computational boundaries
($r_b \rightarrow \infty$), the solutions to these
modified post-Newtonian equations
reduce to the standard post-Newtonian inspiral solutions starting with
circular orbit initial data.  The constants $\sigma_1$ through $\sigma_8$
are to be chosen so as to best reproduce the effects of the truncation
errors and boundary errors from the multiple-orbit simulations of 
binary neutron stars in~\cite{Miller:2003vc}.  To this end, I
define an ``error function'' $\chi{(\sigma_i)}$, which measures the
difference in the orbital separation profile in time
between the 5 numerical relativity simulations NS-A through NS-E
from~\cite{Miller:2003vc} (which I denote as 
$r_{\tt{nr}}^j(t)$,~$j=1,2,3,4,5$) and the solution to the
modified post-Newtonian equations 
Eqs.~\ref{eq:pneom},~\ref{eq:phidotmodel},~and~\ref{eq:ldotmodel} 
(which I denote as
$r_{\tt{pn}}(t,\Delta x,r_b,\sigma_i)$), as
\begin{equation}
\chi({\sigma_i}) = \frac {1}{5} \sum_{j=1}^{5}
\sqrt{ \frac { \int_{0}^{t_f} dt 
   {\left ( r_{\tt{pn}}(t,{(\Delta x)}_j,{(r_b)}_j,
                                                  \sigma_i) -
                                           r_{\tt{nr}}^j(t) \right )}^2 }
             { \int_{0}^{t_f} dt} }
\label{eq:minfunc}
\end{equation}
where ${(\Delta x)}_j$ and ${(r_b)}_j$ are the discretization and
boundary placement parameters used to produce the numerical
relativity result $r_{\tt{nr}}^j(t)$ for each of the five multiple-orbit
numerical relativity simulations of binary neutron stars performed
in~\cite{Miller:2003vc}.  The time integrations in Eq.~\ref{eq:minfunc}
use $t_f = 610 \: m$, which corresponds to roughly $2.5$ orbits of the 
binary system.  A global minimum of $\chi({\sigma^{min}_i})~=~0.093 \: m$ 
is found numerically 
by varying the $\sigma_i$ parameters.  A comparison among the five
numerical relativity simulations NS-A through NS-E 
from~\cite{Miller:2003vc} and
the solutions of the modified post-Newtonian equations of motion
Eqs.~\ref{eq:pneom},~\ref{eq:phidotmodel},~and~\ref{eq:ldotmodel} using the 
$\sigma^{min}_{i}$ parameters that minimize $\chi({\sigma_i})$ is 
shown in Fig.~\ref{fig:numrelpncomp}.

\begin{figure}
\vspace{0.0cm}
\hspace{0.0cm}
\psfig{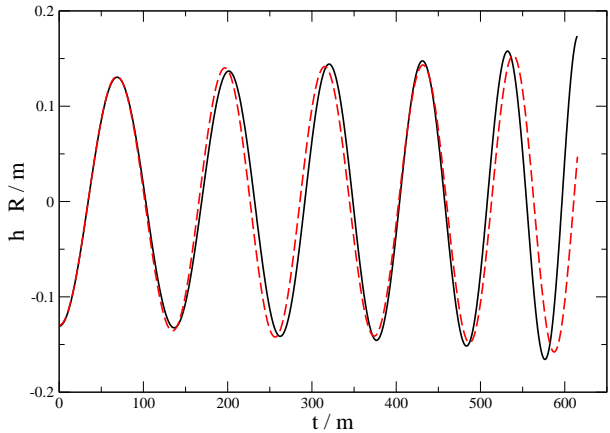}
\caption{Gravitational waveforms (multiplied by the distance to the source,
$R$).  The solid line waveform corresponds to ${h_0}(t)$, which is 
the ``zero error'' waveform corresponding to the 
modified post-Newtonian solution with discretization parameter
$\Delta x = 0$ and boundary placement parameter $r_b = \infty$.
The dashed line waveform is 
$h(t,{\Delta x}_{\tt{NS-A}},{r_b}_{\tt{NS-A}})$,  which corresponds to 
the numerical relativity simulation NS-A (see Fig.~\ref{fig:numrelpncomp}).
The difference between these waveforms (Eq.~\ref{eq:deltah}) corresponds
to $\Delta = 0.054$, which is 5 times larger than our target accuracy
(see Eqs.~\ref{eq:delta}~and~\ref{eq:delta100th}).
}
\vspace{0.0cm}
\label{fig:waveform}
\end{figure}

Using the solutions to the 
modified post-Newtonian equations of motion corresponding
to parameters $\sigma^{min}_{i}$ that minimize $\chi({\sigma_i})$ as
a model for the effects of the truncation and boundary errors in the 
full numerical relativity simulations of orbiting binary
neutron stars in~\cite{Miller:2003vc}, I am now able to gauge the effect
these errors have on the resulting gravitational waveform.  Specifically,
the goal is to find bounds on the discretization parameter $\Delta x$ and 
outer boundary location parameter $r_b$ such that the error in
the produced gravitational waveform satisfies, e.g., Eq.~\ref{eq:delta100th}.
In order to calculate $\Delta \equiv (1/2) 
{ \left \langle \delta h | \delta h \right \rangle} /
      { \left \langle h | h \right \rangle }$,
I take the target gravitational
waveform $h(t)$ to be that determined by the solution to the modified
post-Newtonian equations of motion in the limit as 
$\Delta x \rightarrow 0$ and $r_b \rightarrow \infty$, which is just
the waveform obtained from the ordinary post-Newtonian equations of
motion assuming initial data corresponding to a circular orbit.  This
waveform we denote as ${h_0}(t)$.  The
``error'' in the waveform $\delta h(t)$ induced by the truncation error
and boundary error in the numerical simulations can then be
calculated as 
\begin{equation}
\delta h(t,{\Delta x},{r_b}) = h(t,{\Delta x},{r_b}) - {h_0}(t)
\label{eq:deltah}
\end{equation}
where $h(t,{\Delta x},{r_b})$ is the waveform obtained by the modified
post-Newtonian equations of motion using discretization parameter
$\Delta x$ and boundary placement parameter $r_b$ (and, of course, 
using the $\sigma_i$ parameters that minimizes $\chi(\sigma_i)$).
Fig.~\ref{fig:waveform} is a plot of the target waveform
${h_0}(t)$ and the 
waveform $h(t,{\Delta x}_{\tt{NS-A}},{r_b}_{\tt{NS-A}})$,
which corresponds to the best numerical relativity simulation NS-A (the 
solid line in Fig.~\ref{fig:numrelpncomp}).

\begin{figure}
\vspace{0.0cm}
\hspace{0.0cm}
\psfig{figure=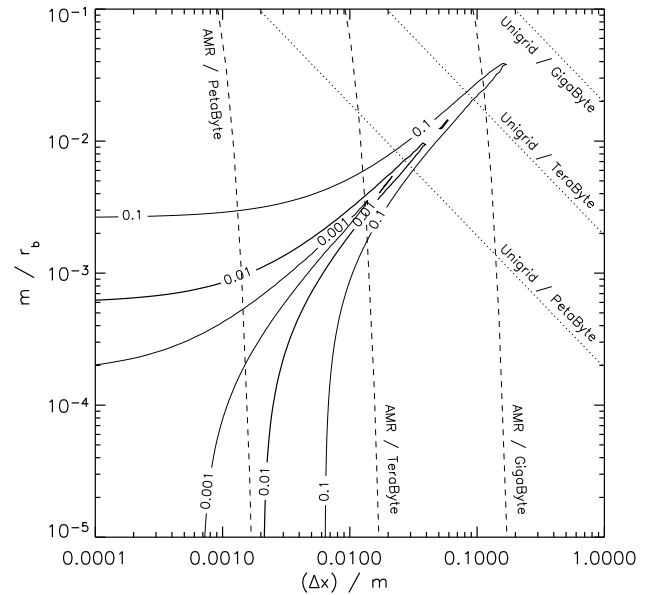,width=8cm}
\caption{Contour plots of the gravitational wave accuracy parameter
$\Delta$ (Eq.~\ref{eq:delta})
as a function of discretization
parameter $\Delta x$ and distance from the center of mass of the
binary to the computational domain boundary $r_b$ for the numerical
relativity simulations of binary neutron stars in~\cite{Miller:2003vc}.
Configurations for
various sized computers with a
unigrid code and an AMR code (nested boxes, with grid refinement ratio
of $2$) is shown for reference.
}
\vspace{0.0cm}
\label{fig:deltacontour}
\end{figure}

Shown in Fig.~\ref{fig:deltacontour} is a contour plot of 
$\Delta$ (Eq.~\ref{eq:delta}) as a function of 
$\Delta x$ and $r_b$.  Contours for $\Delta = 0.1$, $0.01$, and
$0.001$ are shown.  The ``peninsula''-like shape of the contours
in Fig.~\ref{fig:deltacontour} are due to the slightly offsetting effect
of the truncation and boundary errors in the numerical simulations
of~\cite{Miller:2003vc}; larger discretization parameters $\Delta x$ tend
to artificially increase the rate at which angular momentum is lost from the 
binary, while closer outer boundary placements (smaller $r_b$)
tend to have the opposite effect.  For reference, the computational
memory resources for the {\it unigrid} numerical relativity code used
in~\cite{Miller:2003vc} is shown 
in Fig.~\ref{fig:deltacontour}, indicating Gigabyte ($1024^3$ bytes),
Terabyte ($1024^4$ bytes), and Petabyte ($1024^5$ bytes) requirements.
Adaptive mesh refinement (AMR) allows the efficient 
minimization of errors induced by the boundary by permitting the
placement of the boundary of the computational domain farther from
the coalescing binary for a fixed amount of computational resources
(other methods could also reduce boundary errors, such as employing
a Cauchy-characteristic matching code~\cite{Bishop96}
or using a null-approaching
slicing far from the center of mass of the binary~\cite{Husa:2002kk}).
The computational
memory resources for an AMR version of the
code is also shown in Fig.~\ref{fig:deltacontour} (I have assumed that the
finest resolution grid is the size of the compact objects, that
the grid at each level has the same computational volume as
every other grid, and that
the grids are uniformly nested with a refinement ratio of 2).

\begin{figure}
\vspace{0.0cm}
\hspace{0.0cm}
\psfig{figure=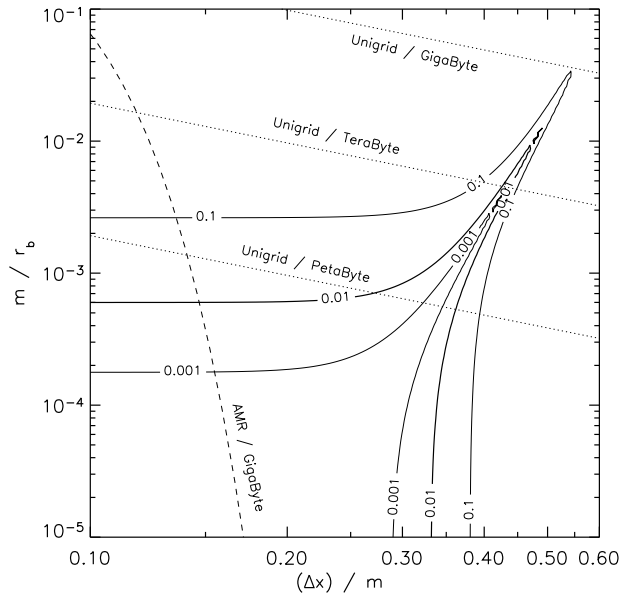,width=8cm}
\caption{Contour plots of $\Delta$ (Eq.~\ref{eq:delta})
as a function of discretization
parameter $\Delta x$ and distance from the center of mass of the
binary to the computational domain boundary $r_b$, assuming an
eighth-order finite difference method is used for the numerical 
relativity simulations of binary neutron stars.  Configurations for various 
sized computers with a
unigrid code and an AMR code is shown for reference.
Note that higher order finite difference methods, couple with AMR, will
be able to robustly obtain the required gravitational 
waveform accuracy $\Delta \leq 0.01$ on Gigabyte computers.
}
\vspace{0.0cm}
\label{fig:deltacontoureight}
\end{figure}

An optimistic reading of Fig.~\ref{fig:deltacontour} implies that 
numerical relativity simulations using a
$10$ Terabyte computer would
be able to attain the target accuracy of 
$\Delta = 0.01$.  However, the reliability and robustness of such a 
calculation would be highly questionable, due to the fact that the
discretization and boundary placement would have to be fine-tuned
to reach the tip of the $\Delta = 0.01$ contour peninsula in
Fig.~\ref{fig:deltacontour}.  In order to obtain a robustly accurate
simulation from which gravitational waveforms could be extracted with
confidence,  it will likely be necessary to, at a minimum, use a target 
resolution ${\Delta x}_{\tt{target}}$ and boundary placement
${r_b}_{\tt{target}}$ such that the gravitational wave accuracy
parameter $\Delta$ (Eq.~\ref{eq:delta}) satisfies 
$\Delta \leq 0.01$ for all ${\Delta x} \leq {\Delta x}_{\tt{target}}$ and
$r_b \geq {r_b}_{\tt{target}}$.  From Fig.~\ref{fig:deltacontour}, 
we see that this minimum target configuration is
at roughly ${\Delta x}_{\tt{target}} \sim 0.002 \: m$ and 
${r_b}_{\tt{target}} \sim 2000 \: m$.  However, this minimum 
target configuration would not be possible with a unigrid code, and would
just barely be possible with an AMR code on a {\it Petabyte} machine, 
although the execution time of such a simulation would render
it highly impractical.
In order to reduce the computational resources required to perform
sufficiently accurate inspiral calculations in numerical relativity,
higher order methods will need to be employed in future calculations.
Possible higher order extensions to the code in~\cite{Miller:2003vc}
include the use of spectral methods (where the truncation error drops
off exponentially with the number of collocation points) or the use
of higher order finite difference methods. 
Fig.~\ref{fig:deltacontoureight} reproduces the results of 
Fig.~\ref{fig:deltacontour} assuming that the truncation error of
the simulation falls off
as ${(\Delta x)}^8$, which is consistent with using an eighth-order
finite difference method.  The increased accuracy of the 
eighth order method allows for a larger discretization parameter
$\Delta x$;  the new minimum target discretization is
${\Delta x}_{\tt{target}} \sim 0.33 \: m$.  As seen in 
Fig.~\ref{fig:deltacontoureight}, a combination of both higher-order
finite difference methods and AMR will yield a robustly accurate 
simulation on Gigabyte scale computers.  Not only will the memory
fingerprint of a sufficiently accurate simulation be relatively
small for an AMR code employing higher-order methods, but
just as importantly, the execution time of a simulation using such a 
code will be considerably smaller than, e.g., the 200,000 CPU hours
required for the NS-A simulation from~\cite{Miller:2003vc} displayed
in Fig.~\ref{fig:numrelpncomp}.

\subsection{generality of results}

Various aspects of the implementation of numerical relativity 
simulations of coalescing binary
compact objects, including details regarding the 
spacetime and hydrodynamics
solvers, boundary conditions, initial data, gauge conditions,
and total evolution times, could
have a large impact on the details of the accuracy studies
presented in sections~\ref{sec:ns1}~and~\ref{sec:ns2}.  For instance,
the implementation of constraint-preserving boundary 
conditions~\cite{Sarbach04a,Kidder04a} in the simulations
presented in~\cite{Miller:2003vc} 
could reduce by a significant amount the errors in the simulation induced
by the outer boundaries.  Of course, any approximation method employed
in the generation of gravitational wave templates used
for signal searches and parameter estimations in gravitational
wave detectors must be validated;  one must directly 
confirm that the approximation is good relative to 
the signal to noise ratio of the detector.  As such, the calculations
presented in sections~\ref{sec:ns1}~and~\ref{sec:ns2} are 
a {\it demonstration} for the case of numerical relativity
simulations of coalescing binary compact objects;  gravitational waveform
results from different numerical relativity codes using 
different methods and/or different implementation techniques must be
validated in a similar way.

\begin{figure}
\vspace{0.0cm}
\hspace{0.0cm}
\psfig{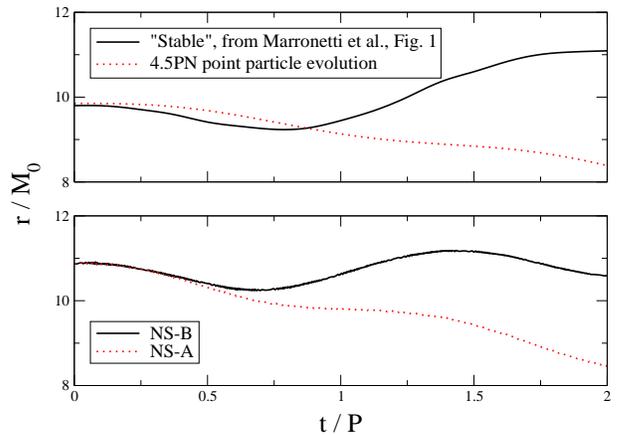}
\caption{ Top panel: the separation of binary neutron stars (normalized by
the total baryonic rest mass of the system $M_0$) simulated 
by the general relativistic hydrodynamics code 
in~\cite{Marronetti04} is plotted as a function of time for two orbital
periods.  For comparison, the solution to the post-Newtonian 
point particle equations
of motion (accurate to order ${(v/c)}^9$) 
for the same masses and initial conditions is shown.  Bottom panel: 
the binary separation from
the general relativistic simulations NS-A and NS-B in~\cite{Miller:2003vc}
(also shown in Fig.~\ref{fig:numrelpncomp} using a different normalization)
is plotted for two orbital periods.  NS-B has a similar outer boundary
placement as that used in the simulations of~\cite{Marronetti04}, but
NS-A has an outer boundary that is located twice as far away from
the binary as that of simulation NS-B.  The similarity of the top
and bottom panels suggests that the unphysical {\it increase} in the 
binary separation of the ``Stable'' simulation in~\cite{Marronetti04} is
due to boundary errors, and that the magnitude of the errors are 
roughly the same for the simulations 
in~\cite{Miller:2003vc} and~\cite{Marronetti04}.
}
\vspace{0.0cm}
\label{fig:marr}
\end{figure}

It is instructive to compare the accuracy requirements
found here with other multiple-orbit binary neutron star simulation 
results, such as those in~\cite{Marronetti04}.
However, the detailed accuracy studies of numerical 
relativity binary simulations in 
sections~\ref{sec:ns1}~and~\ref{sec:ns2} are made possible by repeating
the same simulation (more specifically, using the same initial data) 
many times using a wide variety of discretization parameters and outer
boundary placements, as presented in~\cite{Miller:2003vc}.  
While 
the simulations presented in~\cite{Marronetti04} used several discretization
parameters and boundary placements, they were performed for 
differing initial data corresponding to an array of initial binary 
separations.  Thus, the detailed studies in 
sections~\ref{sec:ns1}~and~\ref{sec:ns2} cannot be repeated using 
the results from the multiple orbit neutron star simulations 
presented in~\cite{Marronetti04}.  However, a casual inspection of the
simulations from~\cite{Marronetti04} confirms that the errors induced by
the boundary are similar to
those of~\cite{Miller:2003vc}.  
In Fig.~\ref{fig:marr}, a simulation of initially 
corotating binary neutron stars from~\cite{Marronetti04} is 
displayed;  the coordinate separation of the neutron stars is plotted
as a function of time over two orbital periods.  The equations of
state used in the simulations of~\cite{Miller:2003vc} and~\cite{Marronetti04}
are identical, and the results from~\cite{Marronetti04} plotted in
Fig.~\ref{fig:marr} use neutron stars that are $7\%$ 
more massive than those used
in~\cite{Miller:2003vc}.  Note that the simulation from~\cite{Marronetti04}
(labeled ``Stable'' in Fig.1 of reference~\cite{Marronetti04})
displayed in the top panel in Fig.~\ref{fig:marr} has an initial
separation of $r_i = 9.85 \: M_0$ (separation values 
in~\cite{Marronetti04} are normalized
by the total baryonic mass $M_0$;  we follow this convention
in Fig.~\ref{fig:marr} and during the discussion here).  After two
orbits, the binary separation has {\it increased} over $10\%$, while 
a post-Newtonian point particle simulation using identical mass, initial
separation, circular orbit initial conditions, and accurate to
order ${(v/c)}^9$ in the post-Newtonian expansion, predicts that the 
separation should instead {\it decrease}
 by $15\%$ during the first two orbits (the
post-Newtonian simulation 
is plotted along side the ``Stable'' simulation
from~\cite{Marronetti04} in the top panel of Fig.~\ref{fig:marr}).
At the very least, it is clear that the separation of the 
neutron stars cannot increase, due to the fact that i) the dissipative effects
of gravitational radiation will cause a decrease in the binary
separation and ii) while the circular orbit initial condition induces a 
slight eccentricity to the orbit of the binary, this initial condition
corresponds to an apastron (maximum separation) point in the dynamical
evolution~\cite{Miller03c}.  To compare with the simulations analyzed
in this paper, the simulations NS-A and NS-B from~\cite{Miller:2003vc}
shown in Fig.~\ref{fig:numrelpncomp} are reproduced in
the lower panel of Fig.~\ref{fig:marr}.  Simulation NS-B has a 
similar outer boundary placement as that used in the ``Stable'' 
simulation from~\cite{Marronetti04}, but NS-A has an outer
boundary that is twice the distance from the binary as compared to NS-B.
The similarities between the top and bottom panels of Fig.~\ref{fig:marr}
suggests that the cause of the 
unphysical increase in the binary 
separation during the first two orbits of the ``Stable'' 
simulation from~\cite{Marronetti04} is the close proximity
of the boundary of the computational domain, and that the 
errors induced by the boundary of the computational domain 
in~\cite{Miller:2003vc} and~\cite{Marronetti04} are of a similar
magnitude.

\section{Conclusions}

Using a criterion for gravitational waveform template accuracy motivated by 
matched filtering and parameter estimation requirements of modern
interferometric gravitational wave detectors, I have calculated the
accuracy required of numerical relativity simulations of coalescing
compact binary systems.  I have calculated
the numerical errors of state-of-the-art numerical
relativity simulations of orbiting binary neutron 
stars~\cite{Miller:2003vc}, and I find these errors to be several orders of 
magnitude larger than the allowed errors determined from
gravitational waveform accuracy considerations.  Using a post-Newtonian
model for the truncation errors and boundary errors in the numerical
simulations of~\cite{Miller:2003vc}, the computational
resources required in order that these simulations 
attain an accuracy needed for reliable gravitational wave extraction
have been calculated.
I find that while mesh refinement technology will provide an 
improvement over the unigrid second-order accurate simulations
of~\cite{Miller:2003vc}, higher order methods will also be required
for a robustly accurate numerical relativity calculation of multiple-orbit
binary coalescence calculations on Terabyte-scaled (or smaller)
digital computers.  

\section{Acknowledgement}

It is a pleasure to thank my colleagues,
both at the Jet Propulsion 
Laboratory and in the California Institute of Technology 
numerical relativity group, for many useful 
discussions and suggestions; special thanks to David Meier for
his many 
useful suggestions regarding this manuscript.
Financial support for this research has been
provided by the
Jet Propulsion Laboratory under contract with the
National Aeronautics and Space Administration.
Computational resource support has been provided by the
JPL Institutional Computing and Information Services, 
the NASA Directorates of Aeronautics Research, Science, Exploration
Systems, and Space Operations, and 
NSF NRAC project MCA02N022.

%\bibliography{./bibtex/references}

\begin{thebibliography}{26}
\expandafter\ifx\csname natexlab\endcsname\relax\def\natexlab#1{#1}\fi
\expandafter\ifx\csname bibnamefont\endcsname\relax
  \def\bibnamefont#1{#1}\fi
\expandafter\ifx\csname bibfnamefont\endcsname\relax
  \def\bibfnamefont#1{#1}\fi
\expandafter\ifx\csname citenamefont\endcsname\relax
  \def\citenamefont#1{#1}\fi
\expandafter\ifx\csname url\endcsname\relax
  \def\url#1{\texttt{#1}}\fi
\expandafter\ifx\csname urlprefix\endcsname\relax\def\urlprefix{URL }\fi
\providecommand{\bibinfo}[2]{#2}
\providecommand{\eprint}[2][]{\url{#2}}

\bibitem[{\citenamefont{Br\"ugmann et~al.}(2004)\citenamefont{Br\"ugmann,
  Tichy, and Jansen}}]{Bruegmann:2003aw}
\bibinfo{author}{\bibfnamefont{B.}~\bibnamefont{Br\"ugmann}},
  \bibinfo{author}{\bibfnamefont{W.}~\bibnamefont{Tichy}}, \bibnamefont{and}
  \bibinfo{author}{\bibfnamefont{N.}~\bibnamefont{Jansen}},
  \bibinfo{journal}{Phys. Rev. Lett.} \textbf{\bibinfo{volume}{92}},
  \bibinfo{pages}{211101} (\bibinfo{year}{2004}), \eprint{gr-qc/0312112}.

\bibitem[{\citenamefont{Alcubierre et~al.}(2004)\citenamefont{Alcubierre,
  Br\"ugmann, Diener, Guzm\'an, Hawke, Hawley, Herrmann, Koppitz, Pollney,
  Seidel et~al.}}]{Alcubierre2003:pre-ISCO-coalescence-times}
\bibinfo{author}{\bibfnamefont{M.}~\bibnamefont{Alcubierre}},
  \bibinfo{author}{\bibfnamefont{B.}~\bibnamefont{Br\"ugmann}},
  \bibinfo{author}{\bibfnamefont{P.}~\bibnamefont{Diener}},
  \bibinfo{author}{\bibfnamefont{F.~S.} \bibnamefont{Guzm\'an}},
  \bibinfo{author}{\bibfnamefont{I.}~\bibnamefont{Hawke}},
  \bibinfo{author}{\bibfnamefont{S.}~\bibnamefont{Hawley}},
  \bibinfo{author}{\bibfnamefont{F.}~\bibnamefont{Herrmann}},
  \bibinfo{author}{\bibfnamefont{M.}~\bibnamefont{Koppitz}},
  \bibinfo{author}{\bibfnamefont{D.}~\bibnamefont{Pollney}},
  \bibinfo{author}{\bibfnamefont{E.}~\bibnamefont{Seidel}},
  \bibnamefont{et~al.}, \bibinfo{journal}{submitted to Physical Review D}
  (\bibinfo{year}{2004}), \eprint{gr-qc/0411149}.

\bibitem[{\citenamefont{Brandt et~al.}(2000)\citenamefont{Brandt, Correll,
  G\'{o}mez, Huq, Laguna, Lehner, Marronetti, Matzner, Neilsen, Pullin
  et~al.}}]{Brandt00}
\bibinfo{author}{\bibfnamefont{S.}~\bibnamefont{Brandt}},
  \bibinfo{author}{\bibfnamefont{R.}~\bibnamefont{Correll}},
  \bibinfo{author}{\bibfnamefont{R.}~\bibnamefont{G\'{o}mez}},
  \bibinfo{author}{\bibfnamefont{M.~F.} \bibnamefont{Huq}},
  \bibinfo{author}{\bibfnamefont{P.}~\bibnamefont{Laguna}},
  \bibinfo{author}{\bibfnamefont{L.}~\bibnamefont{Lehner}},
  \bibinfo{author}{\bibfnamefont{P.}~\bibnamefont{Marronetti}},
  \bibinfo{author}{\bibfnamefont{R.~A.} \bibnamefont{Matzner}},
  \bibinfo{author}{\bibfnamefont{D.}~\bibnamefont{Neilsen}},
  \bibinfo{author}{\bibfnamefont{J.}~\bibnamefont{Pullin}},
  \bibnamefont{et~al.}, \bibinfo{journal}{Phys. Rev. Lett.}
  \textbf{\bibinfo{volume}{85}}, \bibinfo{pages}{5496} (\bibinfo{year}{2000}).

\bibitem[{\citenamefont{Miller et~al.}(2004)\citenamefont{Miller, Gressman, and
  Suen}}]{Miller:2003vc}
\bibinfo{author}{\bibfnamefont{M.}~\bibnamefont{Miller}},
  \bibinfo{author}{\bibfnamefont{P.}~\bibnamefont{Gressman}}, \bibnamefont{and}
  \bibinfo{author}{\bibfnamefont{W.-M.} \bibnamefont{Suen}},
  \bibinfo{journal}{Phys. Rev. D} \textbf{\bibinfo{volume}{69}},
  \bibinfo{pages}{064026} (\bibinfo{year}{2004}), \eprint{gr-qc/0312030}.

\bibitem[{\citenamefont{Shibata and Uryu}(2002)}]{Shibata02a}
\bibinfo{author}{\bibfnamefont{M.}~\bibnamefont{Shibata}} \bibnamefont{and}
  \bibinfo{author}{\bibfnamefont{K.}~\bibnamefont{Uryu}},
  \bibinfo{journal}{Prog. Theor. Phys.} \textbf{\bibinfo{volume}{107}},
  \bibinfo{pages}{265} (\bibinfo{year}{2002}).

\bibitem[{\citenamefont{Marronetti et~al.}(2004)\citenamefont{Marronetti, Duez,
  Shapiro, and Baumgarte}}]{Marronetti04}
\bibinfo{author}{\bibfnamefont{P.}~\bibnamefont{Marronetti}},
  \bibinfo{author}{\bibfnamefont{M.}~\bibnamefont{Duez}},
  \bibinfo{author}{\bibfnamefont{S.}~\bibnamefont{Shapiro}}, \bibnamefont{and}
  \bibinfo{author}{\bibfnamefont{T.}~\bibnamefont{Baumgarte}},
  \bibinfo{journal}{Phys. Rev. Lett.} \textbf{\bibinfo{volume}{92}},
  \bibinfo{pages}{141101} (\bibinfo{year}{2004}),
  \bibinfo{note}{gr-qc/0312036}.

\bibitem[{\citenamefont{\'{E}anna \'{E}.~Flanagan and
  Hughes}(1998{\natexlab{a}})}]{Flanagan97b}
\bibinfo{author}{\bibnamefont{\'{E}anna \'{E}.~Flanagan}} \bibnamefont{and}
  \bibinfo{author}{\bibfnamefont{S.~A.} \bibnamefont{Hughes}},
  \bibinfo{journal}{Phys. Rev. D} \textbf{\bibinfo{volume}{57}},
  \bibinfo{pages}{4566} (\bibinfo{year}{1998}{\natexlab{a}}).

\bibitem[{\citenamefont{Damour and Deruelle}(1981)}]{Damour81}
\bibinfo{author}{\bibfnamefont{T.}~\bibnamefont{Damour}} \bibnamefont{and}
  \bibinfo{author}{\bibfnamefont{N.}~\bibnamefont{Deruelle}},
  \bibinfo{journal}{Phys. Lett.} \textbf{\bibinfo{volume}{87A}},
  \bibinfo{pages}{81} (\bibinfo{year}{1981}).

\bibitem[{\citenamefont{Itoh et~al.}(2001)\citenamefont{Itoh, Futamase, and
  Asada}}]{Itoh01}
\bibinfo{author}{\bibfnamefont{Y.}~\bibnamefont{Itoh}},
  \bibinfo{author}{\bibfnamefont{T.}~\bibnamefont{Futamase}}, \bibnamefont{and}
  \bibinfo{author}{\bibfnamefont{H.}~\bibnamefont{Asada}},
  \bibinfo{journal}{Phys. Rev. D} \textbf{\bibinfo{volume}{63}},
  \bibinfo{pages}{064038} (\bibinfo{year}{2001}).

\bibitem[{\citenamefont{Pati and Will}(2002)}]{Pati02}
\bibinfo{author}{\bibfnamefont{M.~E.} \bibnamefont{Pati}} \bibnamefont{and}
  \bibinfo{author}{\bibfnamefont{C.~M.} \bibnamefont{Will}},
  \bibinfo{journal}{Phys. Rev. D} \textbf{\bibinfo{volume}{65}},
  \bibinfo{pages}{104008} (\bibinfo{year}{2002}).

\bibitem[{\citenamefont{Blanchet et~al.}(1998)\citenamefont{Blanchet, Faye, and
  Ponsot}}]{Blanchet98}
\bibinfo{author}{\bibfnamefont{L.}~\bibnamefont{Blanchet}},
  \bibinfo{author}{\bibfnamefont{G.}~\bibnamefont{Faye}}, \bibnamefont{and}
  \bibinfo{author}{\bibfnamefont{B.}~\bibnamefont{Ponsot}},
  \bibinfo{journal}{Phys. Rev. D} \textbf{\bibinfo{volume}{58}},
  \bibinfo{pages}{124002} (\bibinfo{year}{1998}).

\bibitem[{\citenamefont{Blanchet and Iyer}(2003)}]{Blanchet03}
\bibinfo{author}{\bibfnamefont{L.}~\bibnamefont{Blanchet}} \bibnamefont{and}
  \bibinfo{author}{\bibfnamefont{B.}~\bibnamefont{Iyer}},
  \bibinfo{journal}{Class. Quantum Grav.} \textbf{\bibinfo{volume}{20}},
  \bibinfo{pages}{755} (\bibinfo{year}{2003}), \eprint{gr-qc/0209089}.

\bibitem[{\citenamefont{Gopakumar et~al.}(1997)\citenamefont{Gopakumar, Iyer,
  and Iyer}}]{Gopakumar97}
\bibinfo{author}{\bibfnamefont{A.}~\bibnamefont{Gopakumar}},
  \bibinfo{author}{\bibfnamefont{B.}~\bibnamefont{Iyer}}, \bibnamefont{and}
  \bibinfo{author}{\bibfnamefont{S.}~\bibnamefont{Iyer}},
  \bibinfo{journal}{Phys. Rev. D} \textbf{\bibinfo{volume}{55}},
  \bibinfo{pages}{6030} (\bibinfo{year}{1997}).

\bibitem[{\citenamefont{Miller}(2004)}]{Miller03c}
\bibinfo{author}{\bibfnamefont{M.}~\bibnamefont{Miller}},
  \bibinfo{journal}{Phys. Rev. D} \textbf{\bibinfo{volume}{69}},
  \bibinfo{pages}{124013} (\bibinfo{year}{2004}), \eprint{gr-qc/0305024}.

\bibitem[{\citenamefont{Epstein and Wagoner}(1975)}]{Epstein75}
\bibinfo{author}{\bibfnamefont{R.}~\bibnamefont{Epstein}} \bibnamefont{and}
  \bibinfo{author}{\bibfnamefont{R.~V.} \bibnamefont{Wagoner}},
  \bibinfo{journal}{Astrophys. J.} \textbf{\bibinfo{volume}{197}},
  \bibinfo{pages}{717} (\bibinfo{year}{1975}).

\bibitem[{\citenamefont{Wagoner and Will}(1976)}]{Wagoner76}
\bibinfo{author}{\bibfnamefont{R.~V.} \bibnamefont{Wagoner}} \bibnamefont{and}
  \bibinfo{author}{\bibfnamefont{C.~M.} \bibnamefont{Will}},
  \bibinfo{journal}{Astrophys. J.} \textbf{\bibinfo{volume}{210}},
  \bibinfo{pages}{764} (\bibinfo{year}{1976}).

\bibitem[{\citenamefont{Turner and Will}(1978)}]{Turner78}
\bibinfo{author}{\bibfnamefont{M.}~\bibnamefont{Turner}} \bibnamefont{and}
  \bibinfo{author}{\bibfnamefont{C.~M.} \bibnamefont{Will}},
  \bibinfo{journal}{Astrophys. J.} \textbf{\bibinfo{volume}{220}},
  \bibinfo{pages}{1107} (\bibinfo{year}{1978}).

\bibitem[{\citenamefont{Lincoln and Will}(1990)}]{Lincoln90}
\bibinfo{author}{\bibfnamefont{C.~W.} \bibnamefont{Lincoln}} \bibnamefont{and}
  \bibinfo{author}{\bibfnamefont{C.~M.} \bibnamefont{Will}},
  \bibinfo{journal}{Phys. Rev. D} \textbf{\bibinfo{volume}{42}},
  \bibinfo{pages}{1123} (\bibinfo{year}{1990}).

\bibitem[{\citenamefont{\'{E}anna \'{E}.~Flanagan and
  Hughes}(1998{\natexlab{b}})}]{Flanagan97a}
\bibinfo{author}{\bibnamefont{\'{E}anna \'{E}.~Flanagan}} \bibnamefont{and}
  \bibinfo{author}{\bibfnamefont{S.~A.} \bibnamefont{Hughes}},
  \bibinfo{journal}{Phys. Rev. D} \textbf{\bibinfo{volume}{57}},
  \bibinfo{pages}{4535} (\bibinfo{year}{1998}{\natexlab{b}}).

\bibitem[{\citenamefont{Kalogera et~al.}(2004)\citenamefont{Kalogera, Kim,
  Lorimer, Burgay, D'Amico, Possenti, Manchester, Lyne, Joshi, McLaughlin
  et~al.}}]{Kalogera04}
\bibinfo{author}{\bibfnamefont{V.}~\bibnamefont{Kalogera}},
  \bibinfo{author}{\bibfnamefont{C.}~\bibnamefont{Kim}},
  \bibinfo{author}{\bibfnamefont{D.}~\bibnamefont{Lorimer}},
  \bibinfo{author}{\bibfnamefont{M.}~\bibnamefont{Burgay}},
  \bibinfo{author}{\bibfnamefont{N.}~\bibnamefont{D'Amico}},
  \bibinfo{author}{\bibfnamefont{A.}~\bibnamefont{Possenti}},
  \bibinfo{author}{\bibfnamefont{R.}~\bibnamefont{Manchester}},
  \bibinfo{author}{\bibfnamefont{A.}~\bibnamefont{Lyne}},
  \bibinfo{author}{\bibfnamefont{B.}~\bibnamefont{Joshi}},
  \bibinfo{author}{\bibfnamefont{M.}~\bibnamefont{McLaughlin}},
  \bibnamefont{et~al.}, \bibinfo{journal}{Astrophys. J. Lett}
  \textbf{\bibinfo{volume}{614}}, \bibinfo{pages}{L137} (\bibinfo{year}{2004}).

\bibitem[{\citenamefont{Gustafsson et~al.}(1995)\citenamefont{Gustafsson,
  Kreiss, and Oliger}}]{Gustafsson95}
\bibinfo{author}{\bibfnamefont{B.}~\bibnamefont{Gustafsson}},
  \bibinfo{author}{\bibfnamefont{H.-O.} \bibnamefont{Kreiss}},
  \bibnamefont{and} \bibinfo{author}{\bibfnamefont{J.}~\bibnamefont{Oliger}},
  \emph{\bibinfo{title}{Time dependent problems and difference methods}}
  (\bibinfo{publisher}{Wiley}, \bibinfo{address}{New York},
  \bibinfo{year}{1995}).

\bibitem[{\citenamefont{Hirsch}(1992)}]{Hirsch92}
\bibinfo{author}{\bibfnamefont{C.}~\bibnamefont{Hirsch}},
  \emph{\bibinfo{title}{Numerical Computation of Internal and External Flows}}
  (\bibinfo{publisher}{Wiley-Interscience}, \bibinfo{year}{1992}).

\bibitem[{\citenamefont{Bishop et~al.}(1996)\citenamefont{Bishop, G{\'o}mez,
  Lehner, and Winicour}}]{Bishop96}
\bibinfo{author}{\bibfnamefont{N.~T.} \bibnamefont{Bishop}},
  \bibinfo{author}{\bibfnamefont{R.}~\bibnamefont{G{\'o}mez}},
  \bibinfo{author}{\bibfnamefont{L.}~\bibnamefont{Lehner}}, \bibnamefont{and}
  \bibinfo{author}{\bibfnamefont{J.}~\bibnamefont{Winicour}},
  \bibinfo{journal}{Phys. Rev. D} \textbf{\bibinfo{volume}{{\bf 54}}},
  \bibinfo{pages}{6153} (\bibinfo{year}{1996}).

\bibitem[{\citenamefont{Husa}(2002)}]{Husa:2002kk}
\bibinfo{author}{\bibfnamefont{S.}~\bibnamefont{Husa}}, \bibinfo{journal}{Lect.
  Notes Phys.} \textbf{\bibinfo{volume}{604}}, \bibinfo{pages}{239}
  (\bibinfo{year}{2002}), \eprint{gr-qc/0204043}.

\bibitem[{\citenamefont{Sarbach and Tiglio}(2004)}]{Sarbach04a}
\bibinfo{author}{\bibfnamefont{O.}~\bibnamefont{Sarbach}} \bibnamefont{and}
  \bibinfo{author}{\bibfnamefont{M.}~\bibnamefont{Tiglio}},
  \bibinfo{journal}{submitted to Journal of Hyperbolic Differential Equations}
  (\bibinfo{year}{2004}), \eprint{gr-qc/0412115}.

\bibitem[{\citenamefont{Kidder et~al.}(2004)\citenamefont{Kidder, Lindblom,
  Scheel, Buchman, and Pfeiffer}}]{Kidder04a}
\bibinfo{author}{\bibfnamefont{L.}~\bibnamefont{Kidder}},
  \bibinfo{author}{\bibfnamefont{L.}~\bibnamefont{Lindblom}},
  \bibinfo{author}{\bibfnamefont{M.}~\bibnamefont{Scheel}},
  \bibinfo{author}{\bibfnamefont{L.}~\bibnamefont{Buchman}}, \bibnamefont{and}
  \bibinfo{author}{\bibfnamefont{H.}~\bibnamefont{Pfeiffer}},
  \bibinfo{journal}{submitted to Physical Review D}  (\bibinfo{year}{2004}),
  \eprint{gr-qc/0412116}.

\end{thebibliography}

%%%%%%%%%%%%%%%%%%%%%%%%%%%%%%%%%%%%%%%%%%%%%%%%%%%%%%%%%%%%%%%%%%%%%%%%%

\end{document}